# Hybridized hyperbolic surface phonon polaritons at α-MoO₃ and polar dielectric interfaces

Qing Zhang,[1,*] Qingdong Ou,[2,3,*,†] Guangwei Hu,[1] Jingying Liu,[3] Zhigao Dai,[4] Michael S. Fuhrer,[2,5] Qiaoliang Bao,[6,†] and Cheng-Wei Qiu[1,†]

[1]Department of Electrical and Computer Engineering, National University of Singapore, Singapore 117583, Singapore.

[2]ARC Centre of Excellence in Future Low-Energy Electronics Technologies (FLEET), Monash University, Clayton, Victoria 3800, Australia.

[3]Department of Materials Science and Engineering, Monash University, Clayton, Victoria 3800, Australia

[4]Engineering Research Center of Nano-Geomaterials of Ministry of Education, Faculty of Materials Science and Chemistry, China University of Geosciences, 388 Lumo Road, Wuhan 430074, China.

[5]School of Physics and Astronomy, Monash University, Clayton, Victoria 3800, Australia

[6]Department of Applied Physics, The Hong Kong Polytechnic University, Hung Hom, Kowloon, Hong Kong, China

[*]These authors contributed equally: Qing Zhang, Qingdong Ou

[†]Correspondence to: qingdong.ou@monash.edu, qiaoliang.bao@gmail.com, chengwei.qiu@nus.edu.sg

**Surface phonon polaritons (SPhPs) in polar dielectrics offer new opportunities for infrared nanophotonics due to sub-diffraction confinement with low optical losses. Though the polaritonic field confinement can be significantly improved by modifying the dielectric environment, it is challenging to break the fundamental limits in photon confinement and propagation behavior of SPhP modes. In particular, as SPhPs inherently propagate isotropically in these bulk polar dielectrics, how to collectively realize ultra-large field confinement, in-plane hyperbolicity and unidirectional propagation remains elusive. Here, we report an approach to solve the aforementioned issues of bulk polar dielectric's SPhPs at one go by constructing a heterostructural interface between biaxial van der Waals material (e.g., α-MoO$_3$) and bulk polar dielectric (e.g., SiC, AlN, and GaN). Due to anisotropy-oriented mode couplings at the interface, the hybridized SPhPs with a large confinement factor (>100) show in-plane hyperbolicity that has been switched to the orthogonal direction as compared to that in natural α-MoO$_3$. More interestingly, this proof of concept allows steerable, angle-dependent and unidirectional polariton excitation by suspending α-MoO$_3$ on patterned SiC air cavities. Our finding exemplifies a generalizable framework to manipulate the flow of nano-light and engineer unusual polaritonic responses in many other hybrid systems consisting of van der Waals materials and bulk polar dielectrics.**

## Introduction

Phonon polaritons (PhPs), quasiparticles formed by the coupling of photons with ionic lattice vibrations in polar crystal lattice[1], exhibit long lifetimes and low optical losses, emerging as an important alternative to plasmonic counterparts for sub-diffraction light-matter interactions and infrared nanophotonic applications. Bulk polar crystals, such as SiC, quartz, AlN, GaN, are known to support surface phonon polaritons (SPhPs) at mid- to far-infrared (IR) region[2-4]. However, the light being squeezed in the form of SPhPs suffers from small confinement factors since the momentum of polaritons ($k_p$) is close to that of free-space photon ($k_0$). Besides, the polar crystals are charge neutral, thus limiting their active tunability upon external electrical or magnetic field. Recently, a new approach of placing ultrathin dielectric layers on polar crystals enables ultra-confined dielectric-tailored SPhPs (hereafter termed as *d*-SPhPs) with a large wavevector ($k_p \gg k_0$)[5-10]. Under the large momentum limit, the *d*-SPhPs mode does not "see" much of the outside material (air), but exhibits sub-wavelength modal size within the dielectric slab and very slow group velocity over a large frequency bandwidth[5,11]. For instance, ~190 times squeezed *d*-SPhP has been reported in a heterostructure of atomic-thin MoS$_2$ layer on SiC[6]. Likewise, reversible optical switching of ~70 times squeezed *d*-SPhPs has also been demonstrated by controlling the structural phase of a phase-change material (Ge$_3$Sb$_2$Te$_6$) attached to quartz substrates[5]. The confinement factors are sensitive to the thickness or the index of coating layers, which opens a new avenue for ultra-compact and tunable devices for infrared applications. However, all these *d*-SPhPs are isotropic along the surface, the observation of directional propagation of in-plane hyperbolic *d*-SPhPs in such systems has so far remained elusive. Recently, biaxial van der Waals (vdW) materials (e.g., α-MoO$_3$[12-14] and V$_2$O$_5$[15]) have risen as new candidates for controlling light at the nanoscale due to their ability to support highly confined, low-loss in-plane anisotropic (elliptic or hyperbolic) PhPs, especially the latter allowing an enhanced density of optical states and ray-like directional polariton propagation. Such natural hyperbolicity originates from the structural anisotropy of the vdW crystals, yet being predetermined by the structural definiteness.

In this work, we demonstrate an approach to reconfigure and engineer in-plane hyperbolic response of *d*-SPhPs via anisotropy-oriented mode coupling at the interfaces between biaxial vdW materials (e.g., α-MoO$_3$) and bulk polar dielectrics (e.g., SiC, AlN, and GaN). In the hybrid system, the polar substrate (e.g., SiC) serves as a source of isotropic SPhPs, which are further hybridized by the anisotropic dielectric environment of α-MoO$_3$, thus regenerating highly confined and in-plane hyperbolic *d*-SPhP modes. Remarkably, this reborn hyperbolic dispersion has been switched to the orthogonal direction as compared to that in natural α-MoO$_3$. Moreover, we found that the hybridized *d*-SPhPs can be further steered by suspending α-MoO$_3$ on patterned SiC nano-cavities (air holes with different topologies), revealing angle-dependent and unidirectional polariton excitation at the edge. We also theoretically investigate the possibility to transplant such a concept to many other hybrid systems consisting of ultrathin vdW materials and bulk polar dielectrics.

## Results

To illustrate the reconfiguration of in-plane hyperbolicity of *d*-SPhPs at the heterostructural interface between biaxial α-MoO$_3$ and polar SiC, we first compare it with natural PhPs in α-MoO$_3$ on SiO$_2$, as schematically shown in Fig. 1a. Biaxial α-MoO$_3$ has three Reststrahlen bands (RBs) in the mid-infrared range[14,16], as detailed in Supplementary Section 1. Here we focus on the overlap frequencies between α-MoO$_3$'s second RB (RB-II, $\varepsilon_{100} < 0$, $\varepsilon_{001} > 0$) and SiC's RB from 820 to 972 cm$^{-1}$. A three-layer transfer matrix formalism (see Supplementary Section 2) is derived to trace the dispersions of PhPs and *d*-SPhPs in the above two systems. At RB-II, PhP modes are supported in α-MoO$_3$ along the [100] crystal direction (Fig. 1b), whereas forbidden along the orthogonal [001] direction due to the in-plane hyperbolicity[12,14,16,17]. Thus far, the potential of the dielectric property of α-MoO$_3$ along the [001] direction has not been fully unlocked and utilized. To this regard, here we explore how the anisotropic dielectric environment of α-MoO$_3$ induces dramatic mode coupling and regenerates highly confined *d*-SPhPs on SiC. As shown in Fig. 1c, the hybridized *d*-SPhPs propagate along the [001] direction with a confinement factor exceeding 100, while at the [100] direction, such *d*-SPhP mode is forbidden because both α-MoO$_3$ and SiC have negative permittivities. The hyperbolic wavefront patterns of PhP and hybrid *d*-SPhP modes are investigated and compared by performing Finite-Difference Time-Domain (FDTD) simulations. As shown in Fig. 1d and 1e, the in-plane field distributions clearly show the orthogonal propagation of these two types of polaritons. Upon further inspection of the mode profile of d-SPhPs at

*yz*-cut plane (inset in Fig. 1e, and Supplementary Fig. S4), we find that the mode is primarily confined at the upper and bottom interfaces of α-MoO₃, with the fields extending out to ~500 nm height in *z* direction on either side and only a small part confined in the ultrathin α-MoO3 dielectric slab. The mode profile is similar to that of the *d*-SPhP modes at the interfaces between MoS$_2$/SiC and GST/quartz[5,6]. Since α-MoO₃ performs dielectric behaviors along the [001] direction that do not support any phonon vibrations, it is suggested that this *d*-SPhP mode inherits from the SPhPs in SiC, but manifests directional propagation with in-plane hyperbolic dispersion. Figure 1f and 1g show the corresponding hyperbolic isofrequency contours calculated by the Fourier Transforms (FT) of the simulated $E_z$. The analytical isofrequency dispersions match well with the numerical results (Supplementary Section 3) [18,19]. The hyperbolic open angle transits from [100] direction to [001] direction after replacing the SiO₂ substrate with polar SiC.

Owing to this hybridized orthogonal hyperbolic response of PhPs and *d*-SPhPs (Fig. 1h), we further propose a new concept of steerable polariton excitation by suspending α-MoO₃ on SiC cavities (air holes with different topologies). We use a dodecagon SiC cavity to verify our hypothesis (Fig. 1i). Inside the nano-cavities, the freestanding α-MoO₃ supports natural PhPs with hyperbolic dispersion along the [100] direction (red line in Fig. 1h), while outside the nano-cavities, α-MoO₃/SiC heterostructure supports *d*-SPhPs with orthogonal dispersion alignment (blue lines in Fig. 1h). Unlike the polaritons in graphene and h-BNwhere similar fringe patterns can be observed parallel to the edge with arbitrary orientation as a result of in-plane isotropic dispersion, the edge-tailored *d*-SPhPs in α-MoO₃ or α-MoO₃/SiC manifest angle-dependent selectivity due to the in-plane anisotropic nature. Fig. 1h illustrates how the edge orientation angle influences the propagation of *d*-SPhPs. The wavevector direction is normal to the edge's orientation. We find that the natural PhPs can only be excited when the orientation angles of wavevectors (solid red arrows) are within the open angle of the in-plane hyperbolic isofrequency dispersion, i.e., $\theta < \theta_{open} = \arctan(\sqrt{-\varepsilon_{100}}/\sqrt{\varepsilon_{001}})$, while the hybrid *d*-SPhPs (solid red arrows) are observed when $\theta_{open} < \theta < 90°$. The dashed arrows represent forbidden polariton mode along that direction. Such angle-dependent wavevector alignment allows unidirectional and steerable excitation of PhP and *d*-SPhP modes in the hybrid system consisting of α-MoO₃ and dodecagon SiC cavity. As illustrated in Fig. 1i, due to the different orientation angles of each dodecagon edge, the single side PhPs (red arrows) can only be excited when $\theta < \theta_{open}$, while the *d*-SPhPs (blue arrows) are only observed when $\theta_{open} < \theta < 90°$.

To experimentally observe the in-plane hyperbolic response of *d*-SPhPs, we used scattering-type scanning near-field optical microscopy (s-SNOM, see Methods) to directly visualize the wavefronts of *d*-SPhPs[20,21]. As schematically shown in Fig. 2a, the polaritons can be directly launched by a metallic antenna, which propagate across the α-MoO₃ surface to interfere with the incident field at the tip. Recording the tip-scattered field yields the hyperbolic wavefronts of *d*-SPhPs with fringe spacing $\lambda_p$.

High-quality and large α-MoO$_3$ flakes are mechanically exfoliated from bulk crystals and then transferred onto the SiC substrate. Metallic diamond antennas were patterned and fabricated on the sample surface by direct laser writing photolithography and e-beam evaporation processes (see Fig. 2b). The apex of the diamond antenna is along the [001] crystal direction of α-MoO$_3$, which provides the necessary momentum to excite the *d*-SPhP modes (see Methods and Fig. S5). Figures 2c-e show the near-field amplitude images measured from a 210 nm-thick α-MoO$_3$ sample on SiC at ω = 936 cm$^{-1}$, 930 cm$^{-1}$ and 925 cm$^{-1}$, respectively. It's noted that the hyperbolic polariton propagates along the [001] direction and forms fringes at the end of the metallic antenna apex. The hyperbolic wavefronts, though only one strong fringe, provide solid experimental evidence of in-plane hyperbolic *d*-SPhPs which has not been observed in SiC before[6,22,23]. These real-space observations agree well with the simulated field distributions (|E|) (Fig. 2f-h). The limited fringe numbers can be explained by the reduced figure of merit [$\gamma^{-1}$ = Re($k_p$)/Im($k_p$)] of SiC based SPhP modes (see Supplementary Section 4). Another possible cause is the large fringe spacing in relatively thick α-MoO$_3$ slab (thickness: 210 nm), for instance, the fringe spacing $\lambda_p$~3 μm (Fig. 1f) at ω = 936 cm$^{-1}$. Because *d*-SPhP mode has negative dispersion slope (Fig. 1c), by decreasing illumination frequency to ω = 925 cm$^{-1}$, the *d*-SPhP becomes more confined and more fringes are observed with smaller fringe spacing $\lambda_p$ ~1.5 μm (Fig. 2h). It is noteworthy that both fringe spacing and the confinement depends on the thickness of α-MoO$_3$. In particular, the confinement factor can achieve as high as $\beta$~30 in a 85 nm-thick slab at ω = 930 cm$^{-1}$ (more results shown in Supplementary Fig. S6), which could be further increased by decreasing the excitation frequency and α-MoO$_3$'s thickness. In addition, the open angles ($\theta_{open\_y}$) of *d*-SPhPs are defined between maximum allowed direction of the hyperbolic isofrequency with respect to the α-MoO$_3$'s [001] crystal axis, which are also sensitive to the frequency. By decreasing the frequency from 936 cm$^{-1}$ to 930 cm$^{-1}$, the open angles decrease from 47° to 38°, in good agreement with the theoretical values derived from the equation $\theta_{open\_y} = \arctan(\sqrt{\varepsilon_{001}}/\sqrt{-\varepsilon_{100}})$ (see Supplementary Fig. S7).

As mentioned earlier, another interesting aspect of the heterostructural interface is that it allows to engineer both PhP and *d*-SPhP hyperbolic polaritons in a single system by suspending α-MoO$_3$ on top of tailored SiC hole cavities. To show a proof of concept, we next numerically and experimentally demonstrate the steerable, angle-dependent and unidirectional polariton excitations resulting from different SiC cavities, as depicted in Fig. 3. We should note that the simultaneous or unidirectional excitations of these two polariton modes in the plane are fundamentally different from those in the geometrically defined α-MoO$_3$ cavities in our recent report[24]. As shown in Fig. 3a, the hole cavities in SiC substrate have different topologies, including circular, square, and triangles with different orientation angles (30°, 45° and 60°). For the circular hole cavity (Fig. 3b), PhPs propagate along all possible directions on the flake within the hole, and an almond shape interference pattern is formed. This is similar to that observed in individual α-MoO$_3$ disk cavity[12], which suggests that the hyperbolic dispersion is along [100] direction. By contrast, outside the hole region, the *d*-SPhPs show hyperbolic wavefront propagating along [001] direction. As for a square hole cavity (Fig. 3c), the PhPs fringes are formed only along [100] direction ($\theta = 0°$) inside the cavity, whereas the $d$-SPhPs reflected from the edge produce fringes along [001] direction ($\theta = 90°$) outside the cavity. The experimental near-field images in the bottom panel clearly corroborate our assumption and the corresponding simulated results. It is noted that the $d$-SPhP fringe is brighter than that of PhPs, which further reveals the surface wave nature of $d$-SPhPs (see the mode profile pattern in Fig. S4, and more discussions in Supplementary Section 7).

We further look into the steering of PhPs and $d$-SPhPs at an in-plane edge that is not parallel or perpendicular to the main optical axis of α-MoO$_3$. As shown at the bottom of Fig. 3a, we designed three triangle hole cavities with two legs parallel to the [100] and [001] directions, and the hypotenuse edges oriented with different angles. Here, the orientation of the hypotenuse edge is represented by the angle $\theta$ (Fig. 3a), and the solid (dashed) arrows denote the polariton excited (or not) according to the exaction condition as we discussed in Fig. 1h and 1i. When $\theta = 30° < \theta_{open}$, there is no intersection between the wavevector and $d$-SPhP dispersion (see Fig. S8a), indicating that $d$-SPhP modes along this direction are forbidden (blue dashed arrow) and only the PhP modes can be excited at the hypotenuse edge to form fringes parallel to it (Fig. 3d). In case of $\theta_{open} \sim 45°$, no polariton fringes can be observed at both sides of the hypotenuse edge (Fig. 3e), because the wavevector at this direction is outside the open angles of both two types hyperbolic dispersions. Particularly, as $\theta$ changes to 60°, the intersection point moves into the $d$-SPhP dispersion (blue line in Fig. S8a). Consequently, PhPs are not allowed inside the cavity while the $d$-SPhPs propagate outside the cavity and form fringes parallel to the hypotenuse edge (Fig. 3f). Note that hyperbolic polariton wavefronts are also observed at the top apex of triangle cavities in the experimental near-field images, which further demonstrared the hyperbolic response of $d$-SPhP modes. Here, owing to the small experimental cavity size, the hyperbolic wavefront excited at each triangle apex would affect the fringes parallel to the hypotenuse edge (especially for results in Fig. 3f). By increasing the cavity sizes, we can minimize the hyperbolic wavefront effect and clearly see the unidirectional fringes at the middle of hypotenuse edge (see Supplementary Fig. S8 for more results). The above findings inspired us to selectively excite the PhPs or $d$-SPhPs by rotating the α-MoO$_3$ flake above an in-plane air-SiC hetero-interface, as schematically illustrated in Fig. 3g. In other words, for an individual edge between air and SiC, the left α-MoO$_3$/air side is expected to support natural PhPs, while the right MoO$_3$/SiC side is expected to support hybrid $d$-SPhPs. However, such two types of polaritons cannot be excited at the same time due to their orthogonal dispersion alignment. By rotational adjustment of α-MoO$_3$ layers, the excitation condition of PhPs and $d$-SPhPs can be selectively satisfied, which thus leads to tunable and directional polariton excitation only at one side of the edge between air and SiC. As shown in Fig. 3h, when the rotation angle $\theta=20°$, means that the normal of the edge (excited polariton wavevector) deviates 20° from the [100] direction. This wavevector direction is within the hyperbolic dispersion of natural PhPs, hence leads to unidirectional PhPs excited at the edge and propagate to the left side. In contrast, when the rotation angle changes to $\theta=70°$ (Fig. 3j), the wavevector alignment occurs within the hyperbolic dispersion of *d*-SPhPs, the propagation of fringes flips to the right side.

We finally investigate the possibility to observe similar in-plane hyperbolic *d*-SPhPs at known heterostructural interfaces but in the different frequency range, or in other hybrid systems consisting of biaxial vdW materials and bulk polar dielectrics. For instance, it is interesting to find α-MoO$_3$ on SiO$_2$ supports hyperbolic PhPs along the [001] direction while shifting the excitation frequency to RB-I (e.g., 600-800 cm$^{-1}$), as shown by the dispersion in Fig. 4a, in sharp contrast with the PhPs along [100] direction in the previous reports[12,14,24]. This is because in-plane hyperbolicity occurs when the permittivity signs of the material are different along orthogonal in-plane directions (*ε*$_{100}$*ε*$_{001}$<0), and the open angle of hyperbolic dispersion is along the axis where the permittivity is negative (e.g., *ε*$_{001}$<0 at RB-I). By exploiting a wide range of polar dielectrics in the mid-infrared to terahertz range[2,3], we find polar dielectrics AlN and GaN have overlapping phonon resonance with RB-I band of α-MoO$_3$. As expected, by changing the substrate to these two polar dielectrics, we observed the hybrid modes with similar negative dispersion slopes. Moreover, the dispersions are switched back to the [100] direction (Fig. 4b). To verify this, we performed FDTD simulations to resolve the field distribution at ω = 760 cm$^{-1}$ for 50 nm-thick α-MoO$_3$ on AlN (Fig. 4c), and at ω = 660 cm$^{-1}$ for 50 nm-thick α-MoO$_3$ on GaN (Fig. 4d). Both of them show hyperbolic wavefronts of *d*-SPhPs along the [100] direction. It is interesting to see that hyperbolic wavefronts also appear along the [001] direction, arising from the second-order PhP modes of α-MoO$_3$ with higher light confinement and smaller fringe spacing. Furthermore, we foresee the concept of heterostructural interfaces to be applicable over an extended frequency range by choosing other anisotropic vdW materials (e.g., α-V$_2$O$_5$[15]) or other RB bands (e.g., the RB bands in terahertz region for α-MoO$_3$[25]). This combination of anisotropic materials with polar substrates affords a new platform for engineering polaritonic response and reconfiguring directional polariton devices in broad spectral bands.

## Conclusions

In summary, we demonstrate an efficient approach to engineer ultra-confined and in-plane hyperbolic *d*-SPhP modes in polar dielectrics by covering with a biaxial vdW material α-MoO$_3$. The hybrid *d*-SPhPs, originating from polar dielectrics, are tailored by the anisotropic dielectric environment of α-MoO$_3$ and manifest orthogonal hyperbolic dispersion as compared to that in natural α-MoO$_3$ slabs. Through delicately engineering SiC hole cavities, angle-dependent and unidirectional hyperbolic phonon polaritons (PhPs or *d*-SPhPs) are achieved, which could be further steered by twisting α-MoO$_3$ crystallographic orientation against SiC cavity edge. Future studies could fuse recently emerging topics such as vdW material based metasurfaces[26], and develop polarization-dependent nanophotonic devices, for example, α-MoO$_3$ thin layer on patterned SiC metasurfaces allowing PhP resonance in horizontal polarization while SPhP resonance in the orthogonal polarization. Our approach may also be generalized to other hybrid platforms of anisotropic materials and polar dielectrics, fascinating the research interest in manipulating phonon polaritons in the mid-infrared to terahertz regions.

## Methods

**Sample fabrication.** The α-MoO$_3$ flakes were mechanically exfoliated from bulk crystals synthesized through the chemical vapor deposition method following our previous report[27]. High quality α-MoO$_3$ flakes were then transferred onto undoped 6H-SiC wafers (PAM-XIAMEN Ltd) via a deterministic dry transfer process. The Au antennas were fabricated on the selected α-MoO$_3$ flakes on 6H-SiC substrates by direct laser writing lithography to define the patterns. Electron-beam evaporation was subsequently used to deposit 35 nm-thick Au in a vacuum with a chamber pressure of <6 × 10$^{-6}$ Torr. The apex of the as-fabricated Au antenna was aligned with the [001] direction of α-MoO$_3$ crystals.

**Nanoimaging**. For real-space infrared imaging[28,29], we used a commercial s-SNOM system (Neaspec GmbH) based on a tapping-mode atomic force microscope (AFM). The incident p-polarized infrared light from a tunable CO$_2$ laser was directed to the AFM tip (NanoWorld) and focused into a nanoscale hotspot at the tip apex. The samples were then raster-scanned by the tip at the oscillating frequency and amplitude of ~300 kHz and ~70 nm, respectively. A pseudo-heterodyne interferometer was used to record the tip-scattered field, with subsequent demodulation at the third harmonics (s$_3$, near-field amplitude) of the tapping frequency to reduce the background noise. The hyperbolic wavefronts along the [001] direction of α-MoO$_3$ crystals on 6H-SiC were observed at the apex of Au antenna.

**Numerical simulations**. We used a finite-difference time-domain method based on commercially available software (Lumerical FDTD, 2020) for full-wave simulations. To launch highly confined PhP or *d*-SPhP modes, we use a dipole polarized along the *z* direction, and the distance between the dipole and the uppermost surface of the sample is 150 nm. The boundary condition is a perfectly matched layer. For the simulations in Fig. 1 and Fig. 4, we monitor the real part of $E_z$ at 100 nm on top of the uppermost surface of the sample, which is then used as the input of a Fourier transform to extract the isofrequency dispersion contours. For the simulations in Fig. 2 and Fig. 3, the samples are illustrated by a plane wave with azimuthal polarization in free space to launch the polariton waves at arbitrary directions. The permittivity of the α-MoO$_3$ layer is obtained from refs.14,16, as shown in Supplementary Information Fig. S1. The permittivities of the SiC, AlN and GaN substrates are taken from refs.3,6.

## Acknowledgements

This project was supported by the National Research Foundation, Prime Minister's Office, Singapore under Competitive Research Program Award NRF-CRP22-2019-0006, the National Key Research & Development Program of China (No. 2016YFA0201902), Shenzhen Nanshan District Pilotage Team Program (LHTD20170006), and ARC Centre of Excellence in Future Low-Energy Electronics Technologies (Grant No. CE170100039). This work was performed in part at the Melbourne Centre for Nanofabrication (MCN) in the Victorian Node of the Australian National Fabrication Facility (ANFF).

## Author contributions

Q.Z. and Q.O. contributed equally to this work. Q.Z., Q. O., Q. B. and C.W.Q. conceived the idea. Q.Z. and G.H. developed the theory, and performed the numerical calculations and full-wave simulations. Q.O. designed the experiments and performed the optical measurements. Q.O. fabricated the samples with the help of J.L. and Z.D. All authors have analyzed and discussed the results. Q.Z., Q.O. Q. B. and C. Q. co-wrote the paper with the input of all authors. Q.B and C. Q. supervised the project.

## Data availability

All relevant data is available from the authors.

## Competing financial interests

The authors declare no competing financial interests.

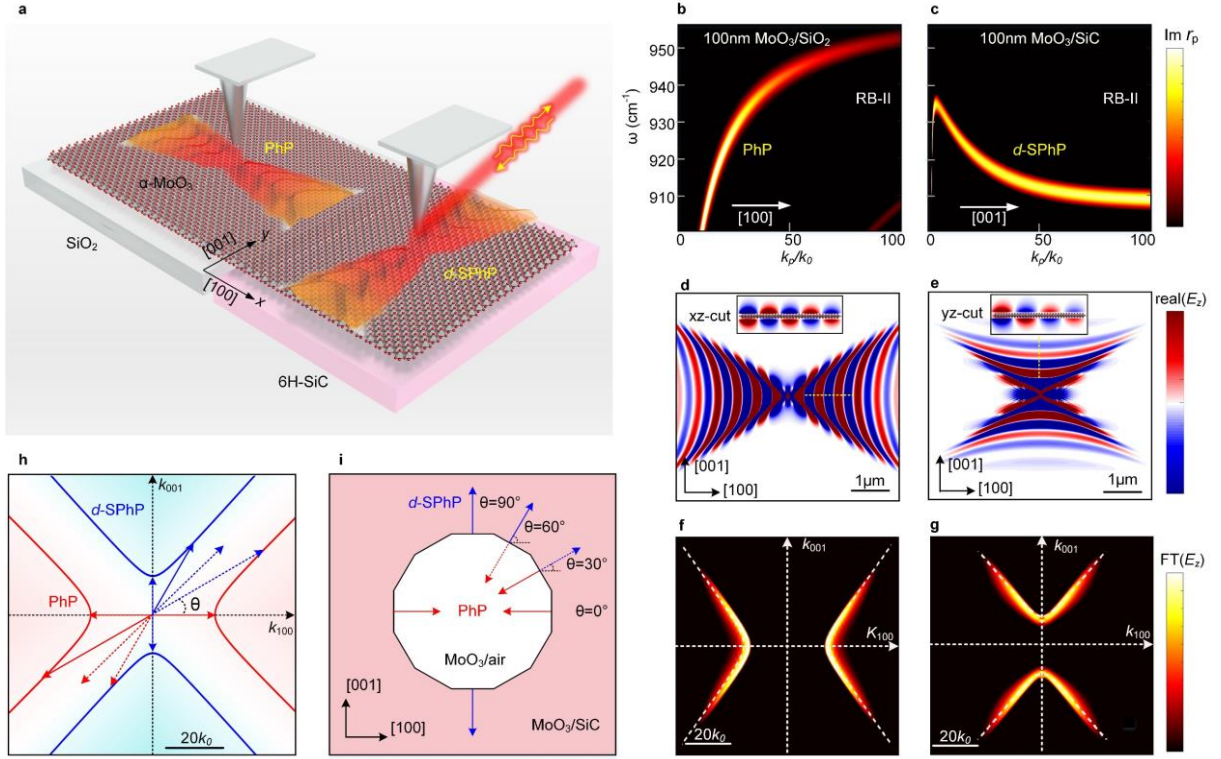

**Figure 1 | Hybridized hyperbolic surface phonon polaritons at the heterostructural interface between biaxial α-MoO₃ and polar SiC. a,** Left panel: schematic of the in-plane hyperbolic PhPs in an α-MoO₃ slab on SiO₂ substrate. The PhPs propagate along [100] crystal direction of α-MoO₃ at RB-II. Right panel: schematic showing the hybrid $d$-SPhPs in heterostructure of α-MoO₃ on polar SiC, in which hyperbolic wavefront propagates along α-MoO₃'s [001] direction. **b, c,** Theoretically calculated (false color plot) dispersions of PhPs and $d$-SPhPs along the α-MoO₃'s [100] or [001] direction, respectively. **d, f,** Numerically simulated field distribution [real($E_z$)] and the corresponding dispersion [Fourier Transforms (FT) of $E_z$] of PhPs in a 100 nm-thick α-MoO₃ slab on SiO₂ substrate. The PhPs are excited by a $z$-polarized dipole at frequency 925 cm$^{-1}$. The white dashed line shows the analytical calculated result. Inset shows the mode profile at the $xz$-cut plane. **e, g,** The field distribution and the FT calculated in-plane dispersion of hybrid $d$-SPhPs. The open angle of hyperbolic dispersion is switched to [001] direction. Inset shows the mode profile at the $yz$-cut plane. **h,** The analytical hyperbolic iso-frequency contours of PhPs (red line) and $d$-SPhPs (blue line) with 100 nm thick α-MoO₃ suspended on air or SiC substrate, respectively. The red arrows denote the wavevector of PhPs at different orientation angle θ with respect to the [100] direction, and the blue arrows denote the wavevector of $d$-SPhPs. While the dashed arrow denotes no polariton mode allowed at this direction. The PhPs can only be excited when θ < θ$_{open}$, while the $d$-SPhPs are observed when θ$_{open}$ < θ < 90°. **i,** Schematic of a dodecagon SiC cavity. By suspending α-MoO₃ above it, the PhPs and $d$-SPhPs can be selectively excited at edges with different orientation angles, and the solid (dashed) arrows denote the polariton excited (or not) according to the wavevector alignment in **h**.

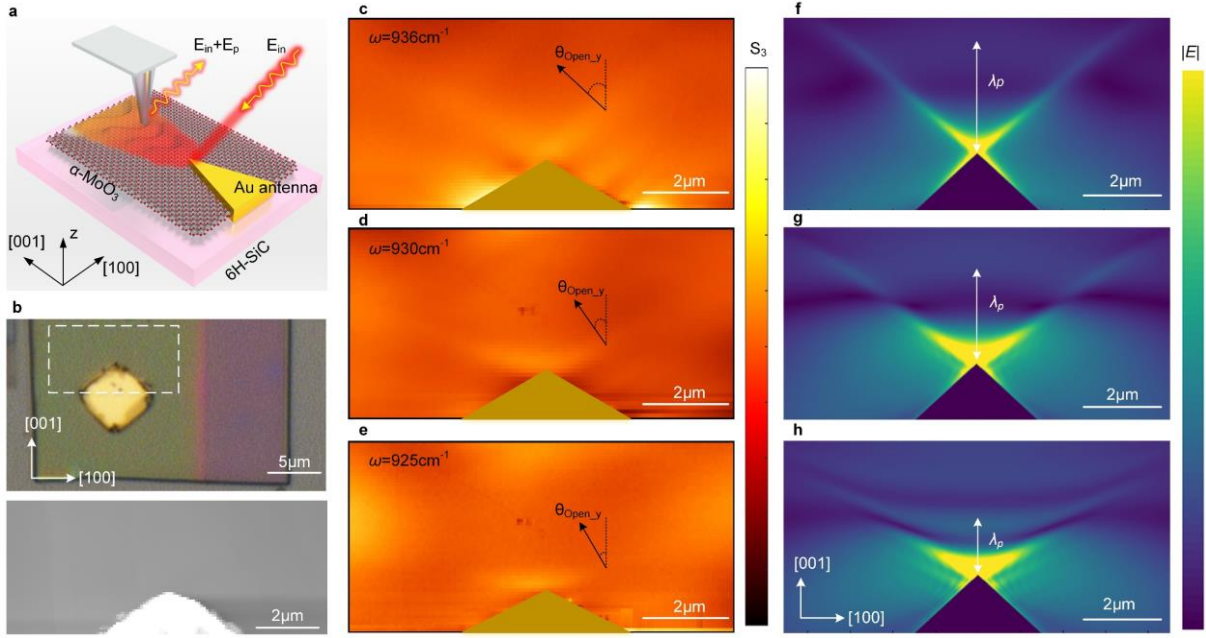

**Figure 2 | Experimental observation of the in-plane hyperbolic *d*-SPhPs. a,** The schematic of the s-SNOM nanoimaging of *d*-SPhPs. Upon infrared illumination ($E_{in}$), the metallic antenna launches the polaritons. The interference field (collected by tip) of incident $E_{in}$ and $E_p$ directly uncover the polariton wavefront. **b,** Optical image (top panel) and AFM image (bottom panel) of α-MoO$_3$ sample on SiC substrate with diamond metallic antenna, and the dashed box denotes the s-SNOM detecting area. **c-e,** Experimentally measured near-field amplitude images ($s_3$) of *d*-SPhPs launched by the diamond metallic antennas. Here, the thickness of α-MoO$_3$ is 210 nm, $\theta_{open\_y}$ denotes the maximum allowed wavevector of the hyperbolic wavefront with respect to the α-MoO$_3$'s [001] crystal axis, which are 47°, 42° and 38° at frequencies 936 cm$^{-1}$, 930 cm$^{-1}$, 925 cm$^{-1}$, respectively. **f-g,** The corresponding numerically simulated field distributions ($|E|$)), $\lambda_p$ represents the wavelength of *d*-SPhPs.

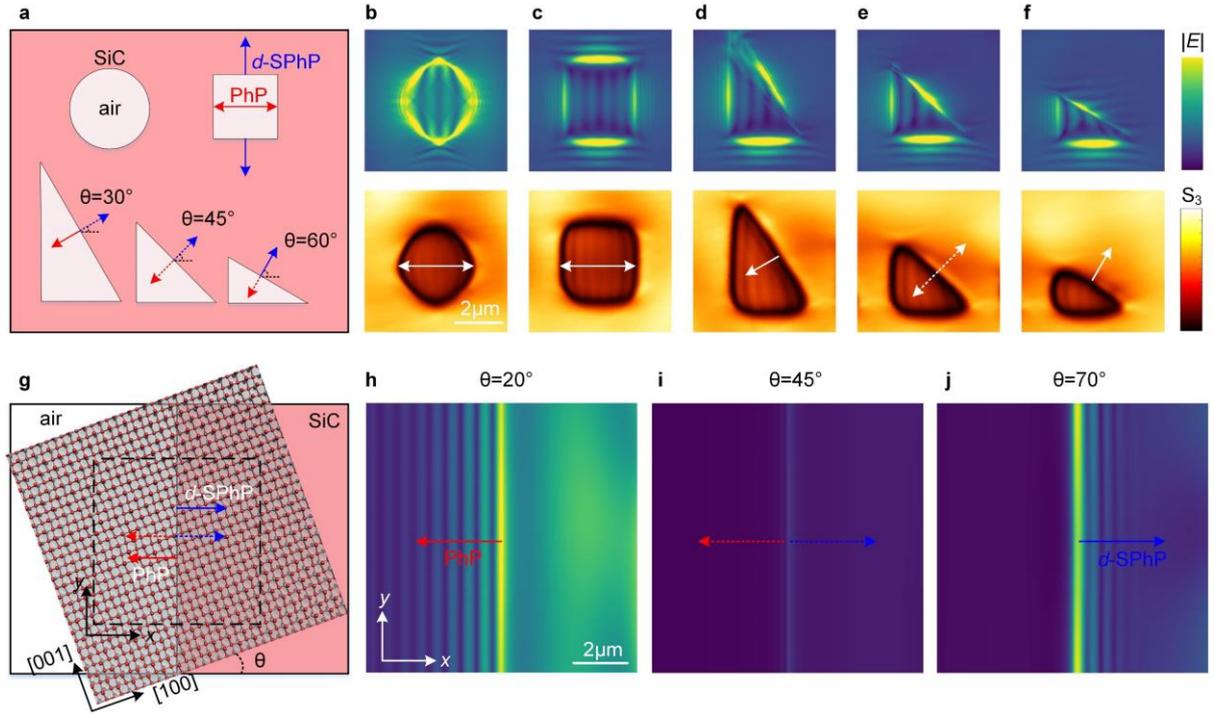

**Figure 3 | Steerable, angle-dependent and unidirectional polariton excitation by controlling the stacking angle between α-MoO₃ and SiC cavities. a,** Schematic of different SiC cavities, such as circular, square, and triangles with different orientation angles (30°, 45°, 60°, respectively). By stacking α-MoO₃ onto the cavities, PhPs (red arrows) and $d$-SPhPs (blue arrows) are selectively excited at the different oriented edges. The solid (dashed) arrows denote the polariton excited (or not). **b-f,** Numerically simulated field distributions ($|E|$, top) and experimentally measured near-field amplitude images ($s_3$, bottom) of PhPs and $d$-SPhPs in different cavities. The experimentally obtained fringe spacing is $\lambda_p/2$ owing to doubled optical path via the collection of the field by the tip, whereas the simulated fringe spacing is $\lambda_p$ as it is propagating wave. The frequency ω = 933 cm⁻¹, and the thickness of α-MoO₃ is $d$=132 nm. The cavity sizes: 3 × 3 μm (b, c), bottom edge 3 μm (d-f). **g,** Schematic of directional excitation of PhPs or $d$-SPhPs by stacking α-MoO₃ onto an air-SiC interface. The rotation angle is defined as the angle θ between the normal of the edge at air-SiC interface with respect to α-MoO₃'s [100] crystal direction. **h,** When θ < θ$_{open}$, it allows for left-side excitation of natural PhP mode. **i,** When θ near to θ$_{open}$, both PhP and $d$-SPhP modes are forbidden. **j,** When θ > θ$_{open}$, only $d$-SPhP mode is excited at the right-side.

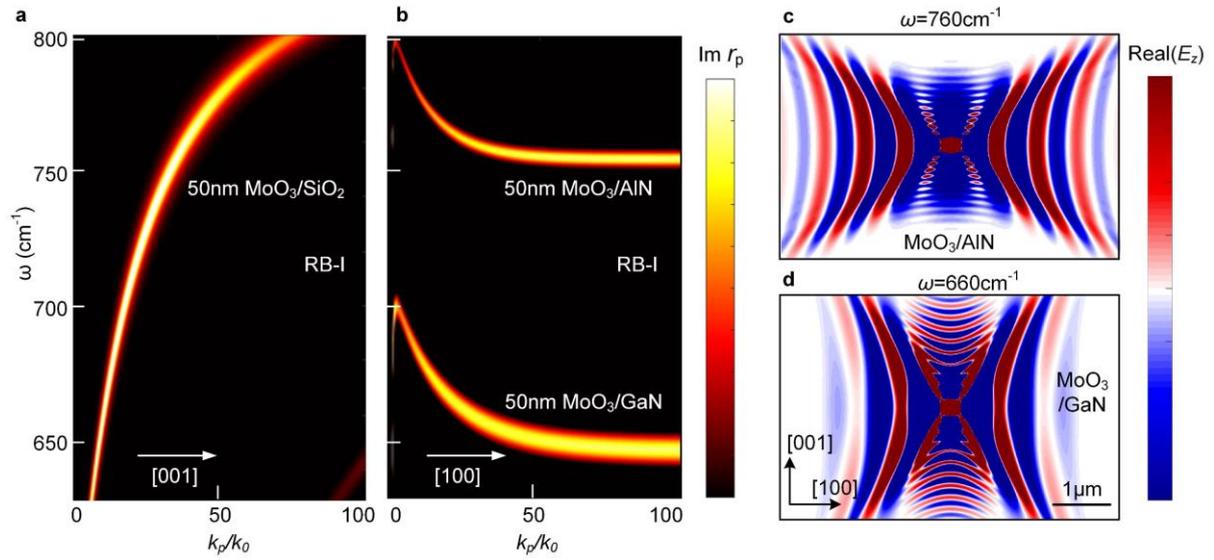

**Figure 4 | In-plane hyperbolic *d*-SPhPs in other hybrid systems of α-MoO₃ and polar dielectrics. a,** The dispersion of PhPs in a 50 nm-thick α-MoO₃ slab on SiO₂ along the [001] crystal direction at RB-I. **b,** The dispersions of *d*-SPhPs in heterostructure of a 50 nm-thick α-MoO₃ slab on AlN (top side) and on GaN (bottom side) along the [100] crystal direction. **c, d,** Numerically simulated field distributions [Real($E_z$)] for α-MoO₃/AlN and α-MoO₃/GaN heterostructure at the frequencies of 760 cm$^{-1}$ and 660 cm$^{-1}$, respectively.